\documentclass[letters,fleqn,usenatbib]{mnras}

\usepackage{mathptmx}
\usepackage{txfonts}

\usepackage[T1]{fontenc}
\usepackage{ae,aecompl}

\usepackage{graphicx}	% Including figure files
\newcommand{\Msun}{$M_{\odot}$}

\newcommand{\MLstar}{$\Upsilon_{\ast}$}
\newcommand{\MLdyn}{$\Upsilon_\mathrm{dyn}$}
\newcommand{\ML}{$\Upsilon$}
\newcommand{\MLref}{$\Upsilon_\mathrm{ref}$}

\newcommand{\A}{$\alpha$}
\newcommand{\GammaB}{$\Gamma_\mathrm{b}$}
\newcommand{\Astar}{$\alpha_{\ast}$}
\newcommand{\Adyn}{$\alpha_\mathrm{dyn}$}

\newcommand{\G}{$\Gamma$}

\newcommand\MgFe{[MgFe]$'$}

\newcommand{\F}{F$_{0.5}$}
\newcommand{\re}{$R_{e}$}

   \title[IMF shape constraints from stellar populations and dynamics]{IMF shape constraints from stellar populations and dynamics  from CALIFA }

   \author[M. Lyubenova et al.]{M. Lyubenova$^1$\thanks{e-mail:lyubenova@astro.rug.nl},
I. Mart\'in-Navarro$^{2,3,4}$,
 G. van de Ven$^5$,
 J. Falc\'on-Barroso$^{2,3}$,
 L. Galbany$^{6,7}$,
  \newauthor
 A. Gallazzi$^{8}$,
 R. Garc\'ia-Benito$^9$,
 R. Gonz\'alez Delgado$^9$,
 B. Husemann$^{10}$,
 F. La Barbera$^{11}$,
   \newauthor
R. A. Marino$^{12}$,
 D. Mast$^{13,14}$,
 J. Mendez-Abreu$^{15}$,
 R.F.P. Peletier$^1$,
 P. S\'anchez-Bl\'azquez$^{16,17}$,
   \newauthor
 S.F. S\'anchez$^{18}$,
 S.C. Trager$^1$,
 R.C.E. van den Bosch$^5$,
 A. Vazdekis$^{2,3}$,
 C.J. Walcher$^{19}$,
   \newauthor
 L. Zhu$^5$,
 S. Zibetti$^{8}$,
 B. Ziegler$^{20}$,
 J. Bland-Hawthorn$^{21}$,
 and  the CALIFA collaboration\\
  $^1$ Kapteyn Astronomical Institute, University of Groningen, Postbus 800, 9700 AV Groningen, the Netherlands\\
  $^2$ Instituto de Astrof\'isica de Canarias, V\'ia L\'actea s/n, La Laguna, Tenerife, Spain \\
  $^3$ Departamento de Astrof\'isica, Universidad de La Laguna, E-38205 La Laguna, Tenerife, Spain\\
  $^4$ University of California Observatories, 1156 High Street, Santa Cruz, CA 95064, USA\\
  $^5$ Max Planck Institute for Astronomy, K\"onigstuhl 17, D-69117 Heidelberg, Germany\\
  $^6$ Millennium Institute of Astrophysics, Chile\\
  $^7$ Departamento de Astronom\'ia, Universidad de Chile, Camino El Observatorio 1515, Las Condes, Santiago, Chile\\
  $^8$ INAF - Osservatorio Astrofisico di Arcetri, Largo Enrico Fermi 5, I-50125 Firenze, Italy\\
  $^9$ Instituto de Astrof\'isica de Andaluc\'ia (CSIC), Glorieta de la Astronom\'ia s/n, Aptdo. 3004, E18080-Granada, Spain\\
  $^{10}$ ESO, Karl-Schwarzscild-Str. 2, D-85748 Garching bei M\"unchen, Germany\\
  $^{11}$ INAF - Osservatorio Astronomico di Capodimonte, Napoli, Italy\\
  $^{12}$ ETH Z\"urich, Institute for Astronomy, Wolfgang-Pauli-Str. 27, 8093 Z\"urich, Switzerland \\
  $^{13}$ Observatorio Astron\'omico, Laprida 854, X5000BGR, C\'ordoba, Argentina\\
  $^{14}$ Consejo de Investigaciones Cient\'{i}ficas y T\'ecnicas de la Rep\'ublica Argentina, Avda. Rivadavia 1917, C1033AAJ, CABA, Argentina\\
  $^{15}$ School of Physics and Astronomy, University of St Andrews, North Haugh, St Andrews, Fife KY16 9SS, Scotland, UK \\
  $^{16}$ Departamento de F\'isica Te\'orica, Universidad Aut\'onoma de Madrid, Cantoblanco, E28049, Spain\\
  $^{17}$ Insistuto de Astrofisica, Pontifica Universidad de Chile, Av. Vicuna Mackenna 4860, 782-0436 Macul, Santiago, Chile \\  
  $^{18}$ Instituto de Astronom\'ia, Universidad Nacional Auton\'oma de M\'exico, A.P. 70-264, 04510, M\'exico, D.F.\\
  $^{19}$ Leibniz-Institut f\"ur Astrophysik Potsdam (AIP), An der Sternwarte 16, D-14482 Potsdam, Germany\\
  $^{20}$ University Vienna, T\"urkenschanzstra{\ss}e 17,1180 Wien, Austria\\
  $^{21}$ Sydney Institute for Astronomy, School of Physics A28, University of Sydney, NSW 2006, Australia
 }

\date{Accepted 2016 June 22; in original form 2016 March 24}

\pubyear{2016}
 
\begin{document}
\label{firstpage}
\pagerange{\pageref{firstpage}--\pageref{lastpage}}
\maketitle

\begin{abstract}
In this letter we describe how we use stellar dynamics information to constrain the shape of the stellar IMF in a sample of 27 early-type galaxies from the CALIFA survey. We obtain dynamical and stellar mass-to-light ratios, \MLdyn\/ and  \MLstar\/, over a homogenous aperture of 0.5~\re\/. We use the constraint  \MLdyn$\ge$\MLstar\/ to test two IMF shapes within the framework of the extended MILES stellar population models. We rule out a single power law IMF shape for 75\% of the galaxies in our sample. Conversely, we find that a double power law IMF shape with a varying high-mass end slope is compatible (within 1$\sigma$) with 95\% of the galaxies. We also show that dynamical and stellar IMF mismatch factors give consistent results for the systematic variation of the IMF in these galaxies. 
\end{abstract}


%%%%%%%%%%%%%%%%%%%%%%%%%%%%%%%%%%%%%%%%%%%%%%%%%%

%%%%%%%%%%%%%%%%% BODY OF PAPER %%%%%%%%%%%%%%%%%%
   
\section{Introduction}
\label{sec:intro}

The stellar initial mass function (IMF) is a fundamental parameter in stellar population theory. Traditionally considered to be universal \citep[e.g.][]{bastian10}, over the last years there is mounting evidence suggesting variations of the IMF, both between and within galaxies. These IMF variations are claimed based on a plethora of methods.  The strength of gravity-sensitive stellar features in the spectra of galaxies implies a larger fraction of low-mass stars with increasing galaxy mass and decreasing radius \citep[e.g.][]{vandokkum10,conroy12,ferreras13,labarbera15,martin-navarro15a}.  Further support for a varying IMF came by constraints from strong gravitational lensing \citep[e.g.][]{treu10,spiniello12,dutton13,smith15} and/or stellar dynamics \citep[e.g.][]{thomas11,cappellari12,tortora13}.

These studies, however, do not always agree on the kind of variation and which process is driving it \citep{smith14}. There are several possible reasons ranging from random sample selections to differences between the adopted stellar population models and how they treat variation of the IMF. The latter has been partially remedied by the comparison of \citet{spiniello15a}.  However, a systematic study to the same data employing a consistent set of assumptions between orthogonal IMF diagnostics is lacking. But before we are able to discuss any consistencies (or the lack thereof), it is worth exploring whether there are particular IMF shapes that are disfavoured by the constraints that the stellar kinematics of galaxies gives us, which are free from any assumptions of the stellar population modelling. This is the aim of the work presented here. We construct dynamical models of a sample of early-type galaxies from the CALIFA survey and use the stringent constraint that the dynamical mass-to-light ratio of a galaxy is an upper limit to the stellar populations mass-to-light ratio.

%######################### end section############

\section{The stellar populations model}
\label{sec:themodel}

\begin{figure}
   \centering
      \includegraphics[width=\hsize]{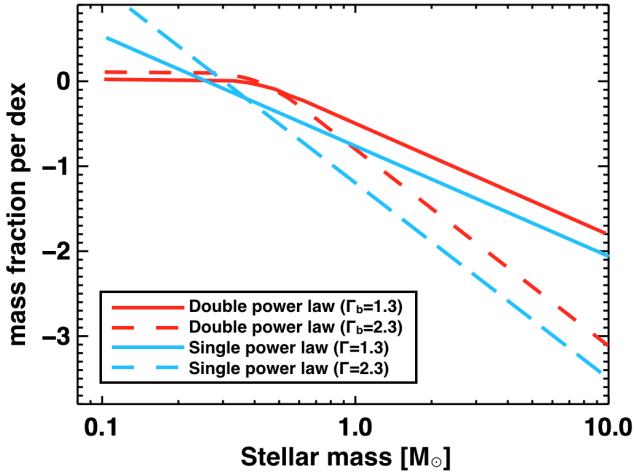}
      \caption{The two IMF shapes studied in this paper.
                          }
         \label{fig:2imf}  
   \end{figure}

For this study, we used the extended version of the MILES models \citep{vazdekis10,vazdekis12}. These models allowed us not only to vary the slope of the IMF, but also its functional form, and thus they were optimally suited for the purpose of this study. We explored two different IMF shapes, namely, single and double power law (see Fig.~\ref{fig:2imf}). The single power law IMF shape, also known as unimodal,  is defined by a single power law, with a logarithmic slope $\Gamma$. Under this parametrisation, a Salpeter IMF \citep{salpeter55} has $\Gamma = 1.35$. The double power law IMF \citep[or bimodal,][]{vazdekis96} is described by two power laws joined by a spline. The only free parameter of that function is the slope of the high-mass end \GammaB\/ (above 0.6~\Msun), and it is tapered for stars with masses below 0.5~\Msun. For \GammaB = 1.3 this double power law IMF is almost indistinguishable from a Kroupa IMF \citep{kroupa01}.

It is important to mention that for relatively old stellar populations (as expected in early-type galaxies), only stars with masses below $\sim$~1 \Msun \, are present. This limits our  stellar populations based IMF analysis to its low-mass end. \citet{labarbera13} have shown that IMF measurements based on stellar population analysis to date are only sensitive to the ratio between stars above and below $\sim$0.5~\Msun \, (\F\/ hereafter). Although the two IMF parameterisations available in the MILES models formally vary the high-mass end of the IMF (through the $\Gamma$ and \GammaB\/ parameters), in practice, the \F\/ is also varied since the models are normalised to a total mass of 1~\Msun.

%######################### end section############

\section{The galaxies and their  mass-to-light ratios}
\label{sec:sample}

We chose our sample of galaxies from the CALIFA survey \citep{sanchez12}. They all reside in the redshift range $0.018 < z < 0.030$ and  have average stellar masses $\sim 10^{12}$~\Msun\/ \citep[Salpeter IMF assumed,][]{gonzalez-delgado14}. 20 of the galaxies were previously analysed by  \citet{martin-navarro15b}. The remaining 7 were observed as part of a dedicated proposal and arefeatured in the  CALIFA Data Release 3 \citep{sanchez16}. The PMAS/PPAK IFU data have all been observed, reduced and analysed in the same way, both in terms of stellar populations and dynamics. All quoted mass-to-light ratios refer to SDSS $r$-band, i.e. \ML\/~$\equiv (M/L)_r$ in solar units.

We chose to explore only the inner 0.5~\re\/ of our sample of galaxies. In this way we minimise the contribution of dark matter to the dynamical mass of the galaxies, which is expected to be increasing at larger radii \citep[e.g.][]{treu04}. We also ensured optimal signal-to-noise ratios to extract consistently dynamics and stellar population parameters.

 \begin{figure*}
   \centering   
   \includegraphics[width=\hsize]{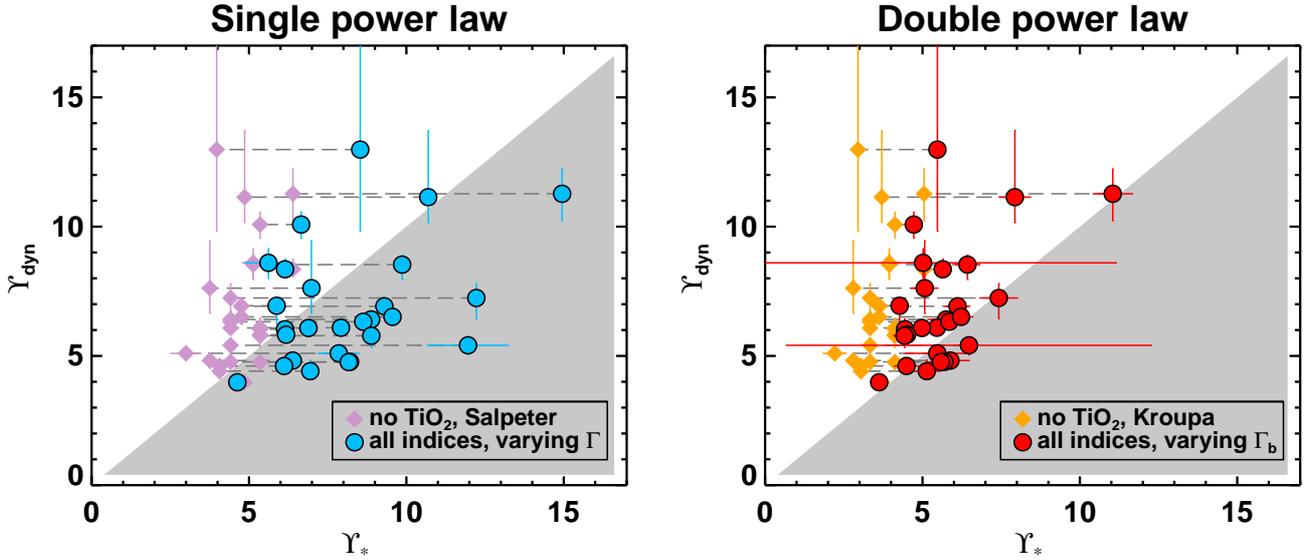}
      \caption{Dynamical versus stellar mass-to-light ratios of our sample of galaxies with the two chosen IMF shapes with varying slopes: single power law ({\it left panel}) and double power law ({\it right panel}). The shaded area represents the "forbidden" region where the requirement of \MLdyn\,$\ge$\,\MLstar\/ is violated. The lilac (Salpeter IMF assumed) and yellow (Kroupa IMF assumed) diamonds represent the \MLstar\/ when the stellar population parameters are derived using the classical diagnostic diagram H$_{\beta_\mathrm{O}}$~---~\MgFe\/. The blue and  red circles denote the galaxies when \MLstar\/ was determined including the stellar gravity-sensitive feature  TiO$_{2_{\mathrm{CALIFA}}}$ and varying slopes of the IMFs. The grey horizontal dashed lines connect the various stellar mass-to-light ratios for the same objects for clarity.
      }
         \label{fig:mldyn_mlstar}
   \end{figure*}

\subsection{Stellar mass-to-light ratios, \MLstar}
\label{sec:mlstar}

To derive the stellar mass-to-light ratios \MLstar\/ we followed the stellar population analysis described in \citet{martin-navarro15b}. In brief, we used the H$_{\beta_\mathrm{O}}$, [MgFe]$^\prime$, and TiO$_{2_{\mathrm{CALIFA}}}$ indices -- an age, metallicity, and IMF slope indicator, respectively -- as these have proved to be the most reliable index combination within the CALIFA wavelength range\footnote{Note that the TiO$_2$ index increases with [$\alpha$/Fe] but decreases with [C/Fe]. Since both $\alpha$-elements and carbon are enhanced in early-type galaxies, the net sensitivity of the TiO$_2$ line to the abundance pattern remains marginal. A more detailed justification of our index selection is given in \citet{martin-navarro15b}}. We measured their values in the spectra integrated over the 0.5~\re\/ elliptical aperture of each galaxy and compared them to the model predictions, convolved to the velocity dispersion of the galaxy. We thus obtained a best-fitting age, metallicity and IMF slope, which were translated into SDSS $r$-band \MLstar\/ predicted by the MILES models. All stellar population parameters are single stellar population (SSP)-equivalent parameters. We consider this choice to be safe following the work of  \citet{labarbera13} and \citet{martin-navarro15b} who showed that non-SSP star formation histories do not significantly alter the stellar population IMF determinations for early-type galaxies.  The stellar $(M/L)_r$, in contrast, does depend on the star formation history, but the old luminosity-weighted ages in our sample ($>$8\,Gyr) ensure a negligible contribution of young stars to the integrated light of our galaxies.
 
We used the 1-$\sigma$ uncertainties of the line-strength measurements, propagated through our analysis, to estimate the errors on the resulting stellar mass-to-light ratios. As our analysis is based on the same set of population models, systematic errors will be common to all models, and so we neglect any additional systematic error component.

For comparison, we also estimated the galaxies' mass-to-light ratios based on their H$_{\beta_\mathrm{O}}$ and \MgFe\/ line strengths only, i.e. without any constraint on the IMF slope,  and assuming the two standard IMF shapes, Kroupa and Salpeter. 

%######################### end section############

\subsection{Dynamical mass-to-light ratios, \MLdyn} 
\label{sec:mldyn}

We derived \MLdyn\/ values of the galaxies by fitting axisymmetric dynamical models to their stellar kinematic maps. In brief, we parametrised the galaxies' stellar surface brightness by applying the multi-Gaussian expansion method \citep[MGE,][]{emsellem94} to their $r$-band images from the $7^{th}$ data release of the Sloan Digital Sky Survey \citep[SDSS,][]{sdss7}. Then we fitted the CALIFA stellar mean velocity and velocity dispersion fields (Falc\'on-Barroso et al., {\it subm.}) with axisymmetric dynamical models, based on a solution of the Jeans equations as implemented by \citet{cappellari08}.  To be consistent with the aperture where the stellar mass-to-light ratio was derived, as explained above, we fitted the stellar kinematics maps only inside the elliptical aperture of 0.5~\re\/. We estimated the inclinations of the galaxies based on their global ellipticity (van de Ven et al., {\it in prep}). We allowed the velocity anisotropy in the meridional plane $\beta_z$ to vary in the range  $(-0.5,1)$. The best-fit based on  $\chi^2$  statistics then yielded our \MLdyn\/ and the corresponding statistical error (typically  $\sim 5\%$). In order to be consistent in our presentation, we did not take into account any systematic errors here, as in the case with \MLstar\/ discussed above. We discuss the effects of these systematic errors on our analysis nn Sect.~\ref{sec:mlstar_mldyn}.

While in principle both the velocity anisotropy and the mass-to-light ratio may vary with radius, we found that this did not significantly improve the fit within 0.5~\re. The strongest gradients in the {\it stellar} \ML\/ are expected in low-mass early-type galaxies \citep{spolaor09,kun10}, where the departure from a standard IMF slope is minimal. Therefore, we do not expect a systematic bias in our measurements due to gradients in the stellar population properties. Whereas we choose to avoid fitting for a particular dark matter distribution, our {\it dynamical} mass-to-light ratio allows for a contribution from dark matter within the explored radius. When we compare with the best-fit stellar mass-to-light ratio, we find small dark matter fractions in agreement with  the $\sim$13\% found previously in such massive early-type galaxies  \citep{cappellari13a}.  The details of our dynamical models and resulting dark matter fractions are described in a forthcoming paper (Lyubenova et al., {\it in prep}).

%######################### end section############

\section{Constraining the shape of the IMF}
\label{sec:mlstar_mldyn}

We used the stringent requirement that the dynamical mass-to-light ratio (\MLdyn\/) should be always greater than or equal to the stellar mass-to-light ratio (\MLstar) to exclude certain types of IMF shapes.

In Fig.~\ref{fig:mldyn_mlstar} we compared the \MLstar\/ and \MLdyn\/ of our sample of galaxies for a selection of IMF shapes. The diamond symbols denote \MLstar\/ when we derived the stellar population parameters using only the H$_{\beta_\mathrm{O}}$ and \MgFe\/ line strengths under the assumption of a Salpeter IMF \citep[lilac diamonds in the {\it left} panel,][]{salpeter55} or a Milky Way-like IMF \citep[yellow symbols in the {\it right} panel,][]{kroupa01}. We refer to these mass-to-light ratios as ``classical''. The blue circles on the {\it left} panel denote our \MLstar\/ with stellar population parameters derived from the complete set of indices (H$_{\beta_\mathrm{O}}$, [MgFe]$^\prime$, and TiO$_{2_{\mathrm{CALIFA}}}$) and a single power law IMF with a varying slope \G\/.  Under this  parametrisation, we find that $0.4  \lesssim \Gamma \lesssim 2.2$ in our sample of galaxies. On the {\it right} panel, the red symbols denote  \MLstar\/ derived in the same way, but with a double power law IMF, whose high-mass end slope \GammaB\/ varies in the range $0.5  \lesssim \Gamma_\mathrm{b} \lesssim 3.1$. The measured ranges of \G\/ and \GammaB\/ values are in agreement with those found by \citet{labarbera16}

While the classical \MLstar\/ (yellow and lilac diamonds) stay within a narrow range, consistent with the narrow range of stellar population properties of our sample of galaxies, \MLdyn\/ shows a much larger spread. If indeed we keep the assumption of a universal IMF, this would be indicative of extremely high dark matter fractions inside the central 0.5 \re\/ for many of the galaxies. The dark matter fractions are less high when using a classical Salpeter IMF instead of a classical Kroupa IMF (lilac versus orange points), but still in disagreement with previous dark matter-fraction derivations \citep[e.g.][]{cappellari13a}. Moreover, a Salpeter IMF does not match the observed TiO$_{2_{\mathrm{CALIFA}}}$ index values. The single power law IMF with varying slope \G\/ (blue symbols on the {\it left} panel) would bring the galaxies with the highest \MLdyn\/ closer to the one-to-one relation with \MLstar\/. However, if such an IMF variation is applied to our complete sample, $\sim$75\% of our galaxies  move to the ``forbidden'' region. Thus we disfavour this single power law IMF shape.

We consider our result to be robust even when including systematic errors. A conservative estimate for these systematics is roughly 20\%  on both \MLstar\/ and \MLdyn\/. The largest uncertainty in our method to estimate \MLstar\/ is driven by the assumptions on the SFH \citep{gallazzi09}. The largest uncertainty on \MLdyn\/ comes from the redshift based distances to the galaxies \citep{jfb11}. However, even when taking into account these conservative error estimates, our result will not be affected.

As evident from the {\it right} panel of Fig.~\ref{fig:mldyn_mlstar}, when we used a double power law IMF with a varying high-mass end slope \GammaB\/ (red symbols) 95\% of the galaxies in our sample are either above the forbidden area or within 1 $\sigma$ of the demarcation line. Our result is consistent with the finding of \citet{labarbera16} who constrain both the shape and the normalisation of the IMF based only on their stellar populations analysis and favour the same double power law IMF.

In the past, galaxies with \MLdyn\/ much higher than SSP-equivalent \MLstar\/ have been explained with having composite star formation histories that would increase the \MLstar\/ compared to the SSP-equivalent ones \citep[e.g.][]{cappellari06,trager08}. However, this explanation was feasible only for the light-weighted properties of low mass galaxies. We consider our result to be robust against composite SFH, as the choice of non-SSP SFH has little effect on the IMF determination in early-type galaxies (see Sect.~\ref{sec:mlstar}).
    
%######################### END SECTION############

\section{Stellar populations and dynamics give consistent results for the IMF variation}
\label{sec:astar_adyn}

\begin{figure}
   \centering
   \includegraphics[width=\hsize]{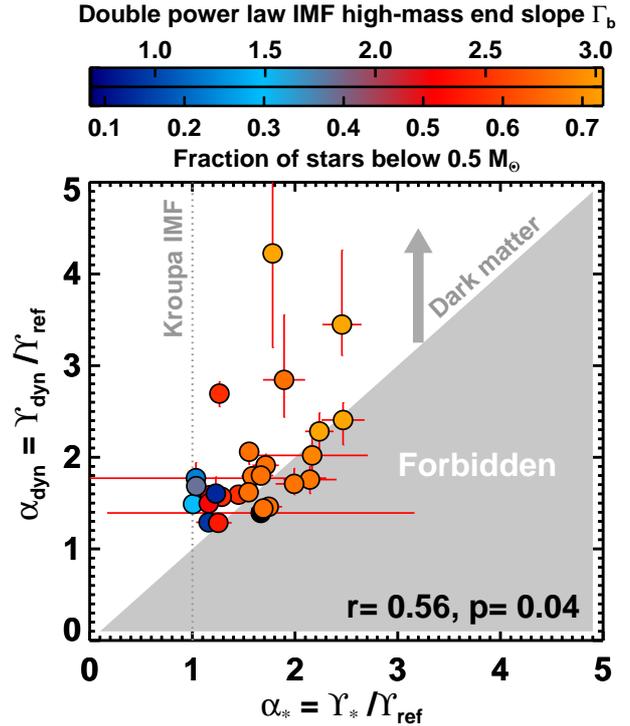}
      \caption{Stellar and dynamical IMF mismatch factors. Their Spearman rank correlation test coefficient $r$ and probability of no correlation $p$ are displayed in the lower right corner. The galaxies are colour coded by their fraction of stars below 0.5~\Msun\/ (i.e. \F), as inferred from their stellar populations. The corresponding slope \GammaB\/ of the high-mass end of a double power law IMF is also given for clarity.
                    }
         \label{fig:alpha_plot}
   \end{figure}

As outlined in Sect.~\ref{sec:intro}, there are several approaches to test the (non)-universality of the IMF. The existence of non-universal IMFs requires at the very least a consistency between stellar population and dynamical estimates of mass-to-light ratios  \citep{smith14}. But a simple correlation between \MLstar\/ and \MLdyn\/, or the lack thereof, cannot be uniquely used to search for variations in the IMF as it can be largely driven by underlying stellar population properties, like age and chemical composition. Indeed, \citet{smith14} has compared literature studies of 34 early-type galaxies and showed that their \MLstar\/ correlate with \MLdyn\/.  However, when he considered the underlying stellar population properties, he found no further correlation on a galaxy-by-galaxy basis. 

We used our sample to test whether stellar populations and dynamics give consistent results about a systematically varying IMF. We first removed the effect on \ML\/ of the underlying stellar population properties, other than the IMF, by normalising the mass-to-light ratios to a reference \MLref\/. This reference is the \ML\/ of a stellar population with the same age and metallicity as the one derived when we allowed the IMF slope to vary but with a fixed IMF shape and normalisation. Our chosen \MLref\/ is that based on a Kroupa IMF (i.e. double power law). This ratio then is often called ``the IMF mismatch factor'', \A.   \Astar$=$\MLstar\/$ /$\MLref\/$=1$ implies that the galaxy has a mass-to-light ratio with the chosen reference IMF (e.g. Milky Way like in our case). \Astar$>1$ indicates departures from this IMF normalisation towards both lower or higher \F\/ (i.e. top- or bottom-heavier IMF) due to the higher fraction of stellar remnants or low mass stars, respectively. \Adyn$=$\MLdyn\/$ /$\MLref\/$=1$ indicates that the galaxy has the chosen reference IMF {\it and no} dark matter content. Thus, \Adyn$>1$ is indicative of either IMF variation or presence of dark matter inside the probed aperture, or both.

  In Fig.~\ref{fig:alpha_plot} we plotted our derived dynamical and stellar IMF mismatch factors. For \Astar\/  we used the mass-to-light ratios coming from the fits with a double power law shape IMF with a varying high-mass end slope. We note that on average our sample has a mean  \Astar$\sim$\Adyn$\sim1.6$. Therefore we conclude, similarly to \citet{smith14}, that the spectroscopic and dynamic claims of a variable IMF agree on average. These average values are remarkably close to the mass scaling factor when one converts from Kroupa to a Salpeter IMF. However, in Sect.~\ref{sec:mlstar_mldyn} we showed that our dynamical mass-to-light ratios allow us to exclude single power-law IMFs,  such as the Salpeter one, for 75\% of our sample, as their stellar mass-to-light ratios become unphysical.

Further, we used the Spearman rank correlation test to probe the correlation between the stellar and the dynamical \A\/, taking into account their uncertainties. The corresponding correlation coefficient is $r=0.56$. This correlation is much stronger than the one found by \citet{smith14}. Moreover, the probability that a correlation between \Astar\/ and \Adyn\/ does not exist is only 4\%. This is in stark contrast with the conclusion of \citet{smith14} that there is no agreement on a case-by-case basis. There are several reasons which might  lead to these differences. \citet{smith14} compares \Astar\/ and \Adyn\/ derived on the same objects but using two very different data sets that cover different extents of the galaxies. \MLdyn\/ is derived from integral field data covering up to 1 \re\/ from the Atlas3D survey \citep{cappellari13a}. On the other hand, \MLstar\/ is derived after fitting the galaxies' spectra integrated over 1/8~\re\/, covering a different and more extensive range of features \citep{conroy12}. However, these aperture differences are unlikely to be the dominant source of scatter.  Actually, approximately half of the compared galaxies lie in the forbidden by dynamics area \citep[see Fig. 1 of][]{smith14}. Therefore, the lack of correlation might in part be caused by unphysical stellar mass-to-light ratios as a result of the adopted IMF shape by \citet{conroy12}.  It is beyond the scope of this letter to investigate the differences in the resulting mass-to-light ratios between the double power law IMF shape that we used here and the three segmented IMF shape of \citet{conroy12}, moreover that the two stellar population analysis methodologies differ significantly. Alternatively, the lack of correlation between \Astar\/ and \Adyn\/ could be due to non-optimal correction of the dynamical mass-to-light ratios for dark matter by \citet{cappellari13a}. This, however, is less likely as these authors tested various dark matter profiles and found similar results. Moreover, we obtained consistent dark matter fractions (see end of Sect.~\ref{sec:mldyn}), despite the different approach in determining this contribution to the total mass budget of the galaxies.

 Almost all of the galaxies in our sample have \Astar\/ and \Adyn\/ values consistently above unity. These variations of the \A\/ values are  larger than the measurement errors and imply inconsistency with a single universal IMF.  For a better illustration of this, we colour-coded the galaxies in Fig.~\ref{fig:alpha_plot} according to their fraction of stars below 0.5~\Msun\/ (\F\/) as  derived from the stellar population analysis. For guidance, we also indicated with a colour bar the inferred high-mass end slope \GammaB\/ of the double power law IMF.  The variation in \Astar\/ directly follows the change in \F\/ as we use the \F\/ to determine \MLstar\/ from the stellar population analysis. However,  \Adyn\/ is independent of the \F\/. Thus the observed correlation between these two -- in other words, galaxies with a higher dynamical IMF mismatch factor have also a higher content of low-mass stars (or bottom-heavier IMF) -- is yet another consistency check for a systematically varying IMF between early-type galaxies.

\section{Conclusions}
\label{sec:conclusions}

The ongoing debate about the exact kind of variation of the stellar initial mass function (IMF) in early-type galaxies poses several challenges in our understanding of galaxy evolution \citep[e.g.][]{martin-navarro16}. Before accepting any of the proposed variations, it is of paramount importance to perform consistency checks by independent methods. In this letter we tested a few particular shapes of the IMF within the framework of the MILES stellar population models \citep{vazdekis10,vazdekis12}. We used a sample of 27 CALIFA galaxies and performed a homogenous analysis of their stellar populations and kinematics. We obtained their stellar and dynamical mass-to-light ratios over the same aperture of 0.5~\re\/ for every galaxy.  Our stellar mass-to-light ratios are determined by the combination of the H$_{\beta_\mathrm{O}}$, [MgFe]$^\prime$, and TiO$_{2_{\mathrm{CALIFA}}}$ indices  -- an age, metallicity, and IMF slope indicator, respectively --  in the integrated spectra of the galaxies. Our dynamical mass-to-light ratios are the result of Jeans axisymmetric dynamical models of the 2-dimensional stellar kinematics. 

After comparing the so derived stellar and dynamical mass-to-light ratios, we find the following results.
\begin{enumerate}
	\item Single power law (unimodal) IMF with a varying slope is excluded for 75\% of the galaxies in our sample.
	\item Conversely, a double power law (bimodal) IMF shape with a varying high-mass end slope is consistent (within 1~$\sigma$) with the dynamical constraints for 95\% of our sample.
	\item Stellar populations and dynamics IMF mismatch factors give consistent results for the IMF variation for our CALIFA sample.
\end{enumerate}

In this letter we have illustrated how valuable the constraints from dynamics can be when determining the shape of the stellar IMF in early-type galaxies. In this  study we chose to limit our analysis to only one stellar population model that allowed us to vary the slope of the IMF and its functional form. The investigation of  other shapes of the IMF, as well as other stellar population models, we leave to a forthcoming paper. Furthermore, this method is applicable not only to the integrated spectra of galaxies but gives promising results when constraining the radial IMF variation and dark matter content of giant early-type galaxies, independent on any assumptions on the dark matter halo profile, as we will show in a forthcoming publication.

\section*{Acknowledgements}
We acknowledge fruitful discussions with Thorsten Lisker and Russell Smith. We thank the reviewer, Richard McDermid, for his valuable comments. Based on observations collected at the Centro Astron\'omico Hispano Alem\'an (CAHA) at Calar Alto, operated jointly by the Max-Planck Institut f\"ur Astronomie and the Instituto de Astrof\'isica de Andaluc\'ia (CSIC).
This paper is based on data obtained by the CALIFA Survey, funded by the Spanish Ministry of Science under grant ICTS-2009-10, and the CAHA.
%Funding and financial support acknowledgements:
 IMN and JFB acknowledge funding from grant AYA2013-48226-C3-1-P from the Spanish Ministry of Economy and Competitiveness (MINECO) and, together with and GvdV, from the FP7 Marie Curie Actions via the ITN DAGAL (grant 289313). 
CJW acknowledges support through the Marie Curie Career Integration Grant 303912.
Support for LG is provided by the Ministry of Economy, Development, and Tourism's Millennium Science Initiative through grant IC120009 awarded to The Millennium Institute of Astrophysics (MAS), and CONICYT through FONDECYT grant 3140566.
RGD acknowledges support from  AyA2014-57490-P.
JMA acknowledges support from the ERC Starting Grant (SEDmorph; P.I. V. Wild).

%-------------------------------------------------------------------
\bibliographystyle{mnras}
\bibliography{/Users/lyubenova/Dropbox/sci/biblio/papers}

% Don't change these lines
\bsp	% typesetting comment
\label{lastpage}
\end{document}